\documentclass[onecolumn,showpacs,preprintnumbers,amsmath,amssymb]{revtex4}
\usepackage{dcolumn}
\usepackage{bm}
\usepackage{epsfig}
\usepackage[usenames]{color}

\begin{document}
\title{Generalized gauge transformation approach to construct dark solitons  of Coupled Nonlinear Schrodinger type  equations}
\author{P. S. Vinayagam$^1$}
\author{R. Radha$^1$}
\email{radha_ramaswamy@yahoo.com}
\author{Vivek M. Vyas$^2$}
\email{vivekmv@imsc.res.in}
\author{K. Porsezian$^3$}
\email{ponzsol@yahoo.com}
\affiliation{$^1$ Centre for Nonlinear Science, PG and Research Dept. of Physics, Govt. College for Women (Autonomous), Kumbakonam 612001, India \\
$^2$Institute of Mathematical
Sciences, Taramani, Chennai 600 113 India\\
$^3$Department of Physics, Pondicherry University,
Pondicherry-605014, India.}

\begin{abstract}
We harnesses the freedom in the celebrated gauge transformation
approach to generate dark solitons of coupled nonlinear
Schr\"odinger (NLS) type equations. The new approach which is
purely algebraic could prove to be very useful, particularly in
the construction of vector dark solitons in the fields of
nonlinear optics, plasma physics  and Bose-Einstein condensates.
We have employed this algebraic method to coupled Gross-
Pitaevskii (GP) and NLS equations and obtained dark solitons.
\end{abstract}
\pacs{03. 75. Lm, 05. 45. Yv}  \maketitle
\section{Introduction}
Eventhough the discovery of solitons in the numerical simulation
of Korteweg de Vries (KdV) equation by Zabusky and Kruskal
\cite{krusal1965} goes back by almost fifty years and several
nonlinear integrable partial differential equations (pdes) like
modified KdV (mKdv) \cite{mkdv}, Sine Gordon(SG) \cite{sg},
nonlinear Schr\"odinger (NLS)\cite{nls} equations have been
identified, construction of integrable dynamical systems and the
associated soliton solution continues to be a challenging problem
even today\cite{ablo}. In the case of NLS equation, soliton type
pulse propagation in the anomalous and the normal dispersion
regimes are called bright and dark solitons,
respectively\cite{hase1,hase2,gpa,darksoliton.nlo}. The
identification of bright solitons in NLS equation arising by
virtue of the subtle balance between the Kerr nonlinearity and
anomalous Group Velocity Dispersion (GVD) has made quite a
turnaround in the fields of optical fibre communications and
responsible for several experimental demonstrations
\cite{hase2,gpa,PSV:PRE,ofc,jrt}. In addition to bright solitons,
it was also reported that optical solitons can also be generated
if GVD is negative (i.e., normal dispersion regime) leading to the
possibility of different type of solitons called dark solitons
\cite{hase2,gpa}. The use of optical solitons opens new prospects
for non-interference transmission since solitons are highly stable
with respect to perturbations caused by fibre nonuniformities and
to external interference\cite{hase2,gpa,jrt}. During the past four
decades or so, it should be mentioned that the NLS equation has
also been used to analyse the dark solitons. Recent studies on the
dark solitons have revealed very interesting properties which may
allow their stable transmission with much less spacing between
solitons when compared with bright solitons. Also, the interaction
effects between two dark solitons are less than the bright
solitons in the presence of fibre loss.  The interaction forces
between two dark solitons are always repulsive, unlike the case of
bright solitons where the interaction forces change according to
their relative phase. The self-induced Raman effect is found to be
more destructive in the case of dark solitons. Use of dark
solitons for high-speed communication systems will remain an
interesting subject for future research \cite{hase2,gpa,dkr,amw}.
It is worth pointing out at this juncture that several analytical
techniques like Hirota bilinear method \cite{hblm,hblm:2}, Darboux
transformation method \cite{mateev,mateev:2}, inverse scattering
method \cite{ist} have been employed to generate bright and dark
solitons.

An iterative approach by gauge transforming the eigenfunctions of
the associated linear eigenvalue problem was developed by Ling Lie
Chau et.al., \cite{llchaw1991} to generate soliton solutions of
integrable nonlinear pdes. Eventhough this method which is quite
similar to Darboux transformation approach is purely algebraic in
nature as it enables one to construct soliton solutions from a
trivial/nontrivial seed solution, the above approach has been
employed only for generating bright solitons for zero seed
solution and kink solitons for non-zero (constant) seed solution
of nonlinear partial differential equations. In this paper, by
choosing a plane wave as the seed solution, we suitably harness
the gauge transformation approach to generate dark solitons of
coupled NLS type equations.

\quad This paper is organized as follows: In section 2, we derive
the algorithm to generate dark soliton solutions by playing around
the freedom in the Gauge transformation method. In section 3, as
an application of the above algorithm, we experiment with coupled
Gross Pitaevskii (GP) and coupled NLS equation and obtain the
corresponding dark soliton solutions.We then discuss the
collisional dynamics of dark solitons.

\section{Modified Gauge transformation approach}
We know that any vector integrable (1+1) dimensional nonlinear
Schr\"odinger equation (in general) can be described by the
celebrated AKNS type linear eigenvalue problem of the following
form

\begin{subequations}
\begin{eqnarray}
\Phi_x &=& U \Phi,\\
\Phi_t &=& V \Phi,
\end{eqnarray}\label{eq:lax}
\end{subequations}
where U and V represent Lax pair governed by 3$\times$3 matrices
and $\Phi = (\phi_1, \phi_2, \phi_3)^T$ is the associated
eigenfunction of the linear eigenvalue problem so that the
compatibility condition

\begin{equation}
U_t-V_x+[U,V]=0\label{eq:zcc}
\end{equation}
generates the desirable integrable vector NLS equation. It should
be mentioned that beginning with non-zero plane wave as the seed
solution is crucial for generating dark solitons. Dark solitons as
is well known, are dips in density profile $|\psi|^{2}$. And
inorder to get such dips, one needs to have a non-zero background
density to begin with which is ensured by the choice of non-zero
plane wave as the seed solution. We now feed the nonzero plane
wave solution into the Lax pair matrices $U$ and $V$ and obtain
the associated linear system.
\begin{subequations}
\begin{eqnarray}
\Phi^{0}_{x} &=& U^{0} \Phi^{0},\\
\Phi^{0}_{t} &=& V^{0} \Phi^{0},
\end{eqnarray}\label{eq:laxnew}
\end{subequations}
where
\begin{equation}
\Phi^{0}=\Phi|_{nonzero\hspace{0.1cm}planewave\hspace{0.1cm}seed}.
\end{equation}

We now gauge transform the eigenfunction $\Phi^{0}$ such that
$\hat{\Phi}= g \Phi^{0}$ where g is a gauge function represented
by a $3 \times 3$ matrix while $\hat{\Phi}$ is an iterated
eigenfunction. The Lax representation in terms of the iterated
eigenfunction is given by
\begin{subequations}
\begin{eqnarray}
\hat{\Phi}_x &+& U_1 \hat{\Phi}=0,\\
\hat{\Phi}_t &+& V_1 \hat{\Phi}=0,
\end{eqnarray}\label{eq:evpnls}
\end{subequations}
so that
\begin{eqnarray}
U_{1}=g U^{0} g^{-1}+g_x g^{-1}\notag\\
V_{1}=g V^{0} g^{-1}+g_t g^{-1}\notag
\end{eqnarray}
Now, the gauge function $g(x,t)$, can be chosen in such a way that
it represents the solution of the associated Riemann problem and
it is meromorphic in the complex $\zeta$ plane as
\begin{equation}
g(x,t;\zeta)=\left[1+\frac{\zeta_1-\bar{\zeta}_1}{\zeta-\zeta_1}P(x,t)
\right]\cdot J\label{gmat}.
\end{equation}
In the above equation, $\zeta$ is the eigenvalue parameter while
$\zeta_1$ and $\bar{\zeta}_1$ represent arbitrary complex
parameters and $P$ is a 3 $\times$ 3 projection matrix. It should
be noted that for scalar NLS type equations, $J=\sigma_{3}$
(Pauli's spin matrix) while for vector NLS type equation
\begin{equation}
J=\left(%
\begin{array}{ccc}
  1 & 0 & 0 \\
  0 & -1 & 0 \\
  0 & 0 & -1 \\
\end{array}%
\right).
\end{equation}
It should be mentioned that, for scalar NLS equations, the gauge
function $g(x,t)$, $U$, $V$ and the projection matrix $P$ are
(2$\times$ 2) in nature and  $\phi$ is a (2 $\times$ 1) column
vector. The inverse of matrix $g$ is given by
\begin{equation}
g^{-1}(x,t;\zeta)=J
\cdot\left[1-\frac{\zeta_{1}-\bar{\zeta}_1}{\zeta-\bar{\zeta}_1}P(x,t)
\right].
\end{equation}
Now, choosing $\bar{\zeta}_1 = \zeta_1^*$, the projection matrix
can be determined by solving the following set of partial
differential equations
\begin{subequations}
\begin{eqnarray}
 P_x =(1-P)J U^{(0)}(\bar{\zeta}_1)J P-PJ U^{(0)} (\zeta_1)J(1-P),\\
 P_t =(1-P)J V^{(0)}(\bar{\zeta}_1)J P-PJ
 V^{(0)}(\zeta_1)J(1-P),
\end{eqnarray}\label{pro}
\end{subequations}
It should be mentioned that equations.(\ref{pro}) are identical to
the one proposed by Ling Lie Chau et.al.,\cite{llchaw1991} in the
gauge transformation approach with a difference that $P, J,
U^{(0)}$ and $V^{(0)}$ are $(2\times 2)$ matrices for scalar
nonlinear pdes while one has to work with $(3 \times 3)$ matrices
for vector nonlinear pdes. One can then relate the projection
matrix $P(x,t)$ with the vacuum eigenfunction $\Phi^{(0)}(x,t)$ by
\begin{equation}
P=J\cdot \tilde{P}\cdot J,\\
\end{equation}
where
\begin{equation}
\tilde{P}= \frac{M^{(1)}}{\rm{Trace} [M^{(1)}]},
\end{equation}
and
\begin{equation}
M^{(1)}=\Phi^{(0)}(x,t,\bar{\zeta}_1)\cdot\hat{m}^{(1)}\cdot\Phi^{(0)}
(x,t,\zeta_1)^{-1}.\label{mmat}
\end{equation}
\begin{equation}
\Phi^{(0)}(x,t,\zeta_1)=\Phi(x,t,\zeta_1)|_{nonzero\hspace{0.1cm}planewave\hspace{0.1cm}seed}.
\end{equation}
In the above equation, $\hat{m}^{(1)}$ is a 3$\times$3 arbitrary
matrix of the form
\begin{equation}
\hat{m}^{(1)}=\left(
\begin{array}{ccc}
e^{2\delta _{1}}\sqrt{2} & \varepsilon _{1}^{(1)}e^{2i(\chi
_{1}+\xi _{1})}
& \varepsilon _{2}^{(1)}e^{2i(\chi _{1}+\xi _{2})} \\
\varepsilon _{1}^{\ast (1)}e^{-2i(\chi _{1}+\xi _{1})} &
e^{-2\delta _{1}}/
\sqrt{2} & 0 \\
\varepsilon _{2}^{\ast (1)}e^{-2i(\chi _{1}+\xi _{2})} & 0 &
e^{-2\delta _{1}}/\sqrt{2} \\ &  &
\end{array}
\right) ,
\end{equation}
where $\chi_{1},\delta_{1},\xi_{1},\xi_{2},$ are arbitrary
functions of (x,t) and their choice is governed by the dispersion
relation of the associated nonlinear pdes while $\varepsilon
_{1}^{(1)}$ and $\varepsilon _{2}^{(1)}$ are the coupling
parameters. It should be noted that the structure of
$\hat{m}^{(1)}$ matrix determines the nature (bright or dark) of
soliton solutions of vector NLS type equations.

Hence, one can write down the dark soliton solution as,
\begin{eqnarray}
\psi_{1}^{(1)} &=&\psi_{1}^{(0)}-2i(\zeta _{1}-\bar{\zeta _{1}})\tilde{P}_{12},\label{darkgenone} \\
\psi_{2}^{(1)}&=&\psi_{2}^{(0)}-2i(\zeta_{1}-\bar{\zeta_{1}})\tilde{P}_{13}.\label{darkgentwo}
\end{eqnarray}
where $\psi_i^{(0)}$(i=1,2) represents the seed solution while
$\psi_{i}^{(1)}$(i=1,2) denotes the iterated dark soliton
solutions of the corresponding coupled NLS type equation and
\begin{equation}
\tilde{P}_{12}=\frac{M^{1}_{12}}{M^{1}_{11}+M^{1}_{22}+M^{1}_{33}}\notag
\end{equation}
\begin{equation}
\tilde{P}_{13}=\frac{M^{1}_{13}}{M^{1}_{11}+M^{1}_{22}+M^{1}_{33}}\notag
\end{equation}
The modified gauge transformation approach can be extended to
generate multidark soliton solutions. For example, the general
form of `N'th dark soliton solution can be written as
\begin{subequations}
\begin{eqnarray}
\psi_1^{(N)} =\psi_1^{(N-1)}-2i (\zeta_N - \bar{\zeta_N})\frac{\tilde{P}_{12}}{R} ,\\
\psi_2^{(N)} =\psi_2^{(N-1)}-2i (\zeta_N -
\bar{\zeta_N})\frac{\tilde{P}_{13}}{R},
\end{eqnarray}\label{eq:twosol}
\end{subequations}
where $\tilde{P}_{12}$ and $\tilde{P}_{13}$ are given by
\begin{eqnarray}
\tilde{P}^{N-1}_{12}=&-&[M_{12}^{(N-1)}((\tau+\gamma M_{11}^{(N-1)})M_{11}^{(N)}+\gamma(M_{12}^{(N-1)}\nonumber\\
&&M_{21}^{(N)}\gamma^*/\tau^2+M_{13}^{(N-1)}M_{31}^{(N)}))+M_{32}^{(N-1)}\nonumber\\
&&((\tau+\gamma M_{11}^{(N-1)})M_{13}^{(N)}+ \gamma(M_{12}^{(N-1)}M_{23}^{(N)} \nonumber\\
&+& M_{13}^{(N-1)}M_{33}^{(N)}))\gamma^*/\tau^2 +((\tau+ \gamma M_{11}^{(N-1)})M_{12}^{(N)} \nonumber\\
&+&\gamma (M_{12}^{(N-1)} M_{22}^{(N)}+M_{13}^{(N-1)}M_{32}^{(N)}))\nonumber\\
&&(\tau + M_{22}^{(N-1)}\gamma*)/\tau^2],\nonumber\\
\tilde{P}^{N-1}_{13}=&-&[M_{13}^{(N-1)}((\tau+\gamma M_{11}^{(N-1)})M_{11}^{(N)}+ \gamma(M_{12}^{(N-1)}\nonumber\\
&&M_{21}^{(N)}+M_{13}^{(N-1)}M_{31}^{(N)}))\gamma^*)/\tau^2+M_{23}^{(N-1)}\nonumber \\
&&((\tau+\gamma M_{11}^{(N-1)})M_{12}^{(N)}+\gamma(M_{12}^{(N-1)}M_{22}^{(N)}\nonumber\\
&+&M_{13}^{(N-1)}M_{32}^{(N)}))\gamma^*/\tau^2+((\tau +\gamma
M_{11}^{(N-1)})M_{13}^{(N)}\nonumber\\
&+&\gamma(M_{12}^{(N-1)}M_{23}^{(N)}+M_{13}^{(N-1)}M_{33}^{(N)}))\nonumber\\
&&(\tau+M_{33}^{(N-1)}\gamma^*)/\tau^2],\nonumber
\end{eqnarray}
and
\begin{eqnarray}
\tau &=& M_{11}^{(N-1)}+M_{22}^{(N-1)}+M_{33}^{(N-1)},\qquad\gamma = \frac{\zeta_1 - \bar{\zeta_1}}{\zeta_2 -\zeta_1},\nonumber\\
R&=&\tilde{P}^{N-1}_{11}+\tilde{P}^{N-1}_{22}+\tilde{P}^{N-1}_{33},\qquad\gamma^*
=-\frac{\zeta_1 - \bar{\zeta_1}}{\zeta_2 -\bar{\zeta_1}},\nonumber
\end{eqnarray}
with
\begin{eqnarray}
\tilde{P}^{N-1}_{11}&=&M_{21}^{(N-1)}((\tau+\gamma
M_{11}^{(N-1)})M_{12}^{(N)}+\gamma(M_{12}^{(N-1)}M_{22}^{(N)}\notag\\
&+&M_{13}^{(N-1)}M_{32}^{(N)}))\gamma^*/\tau^2+M_{31}^{(N-1)}((\tau+\gamma
M_{11}^{(N-1)})\notag\\
&&M_{23}^{(N)}+M_{13}^{(N-1)}M_{33}^{(N)}))\gamma^*/\tau^2 M_{13}^{(N)}+\gamma(M_{12}^{(N-1)}\nonumber\\
&&+((\tau+\gamma M_{11}^{(N-1)})M_{11}^{(N)}+\gamma(M_{12}^{(N-1)}M_{21}^{(N)}+ \nonumber\\
&&M_{13}^{(1)}M_{31}^{(N)}))(\tau+M_{11}^{(N-1)}\gamma^*)/\tau^2,\nonumber\\
\tilde{P}^{N-1}_{22}&=&M_{12}^{(N-1)}(\gamma M_{11}^{(N)}M_{21}^{(N-1)}+M_{21}^{(N)}(\tau+\gamma M_{22}^{(N-1)}) \nonumber\\
&+&\gamma M_{23}^{(N-1)} M_{31}^{(N)})\gamma^*/\tau^2+ M_{32}^{(N-1)}(\gamma M_{13}^{(N)} M_{21}^{(N-1)}  \nonumber\\
&+&(\tau + \gamma M_{22}^{(N-1)}) M_{23}^{(N)}+\gamma M_{23}^{(N-1)}M_{33}^{(N)})\gamma^*/\tau^2 \nonumber\\
&+&(\gamma M_{12}^{(N)}M_{21}^{(N-1)}+(\tau+\gamma M_{22}^{(N-1)}) M_{22}^{(N)}+\gamma M_{23}^{(N-1)}\nonumber\\
&&M_{32}^{(N)})(\tau+M_{22}^{(N-1)}\gamma^*))/\tau^2\notag\\
\tilde{P}^{N-1}_{33}&=&M_{13}^{(N-1)}(\gamma M_{11}^{(N)} M_{31}^{(N-1)} + \gamma M_{21}^{(N)}M_{32}^{(N-1)}+M_{31}^{(N)} \nonumber\\
&&(\tau+\gamma M_{33}^{(N-1)}))\gamma^*/\tau^2+ M_{23}^{(N-1)}(\gamma M_{12}^{(N)}M_{31}^{(N-1)}\nonumber\\
&+& \gamma M_{22}^{(N)} M_{32}^{(N-1)}+M_{32}^{(N)}(\tau + \gamma
M_{33}^{(N-1)}))\gamma^*/\tau^2\notag\\
&+&(\gamma M_{13}^{(N)} M_{31}^{(N-1)}+\gamma
M_{23}^{(N)}M_{32}^{(N-1)}+ (\tau \nonumber\\
&+&\gamma M_{33}^{(N-1)})
M_{33}^{(N)})(\tau+M_{33}^{(N-1)}\gamma^*)/\tau^2,\nonumber
\end{eqnarray}

\begin{eqnarray}
M_{11}^{(N)}&=&e^{-\theta_N}\sqrt{2};\quad\nonumber
M_{12}^{(N)}=e^{-i\xi_N+\chi
_{1}}\varepsilon_1^{(N)};\quad\nonumber
M_{13}^{(N)}=e^{-i\xi_N+\chi _{1}}\varepsilon_2^{(j)};\nonumber\\
M_{21}^{(N)}&=&e^{i\xi_N+\chi
_{1}}\varepsilon_1^{*(N)};\quad\nonumber
M_{22}^{(N)}=e^{\theta_N}/\sqrt{2};\quad\nonumber
M_{23}^{(N)}=0;\nonumber\\
M_{31}^{(N)}&=&e^{i\xi_N+\chi
_{1}}\varepsilon_2^{*(N)};\quad\nonumber
M_{32}^{(N)}=0;\quad\nonumber
M_{33}^{(N)}=e^{\theta_j}/\sqrt{2},\nonumber
\end{eqnarray}

To generate dark solitons for the scalar  NLS type equations, one
can feed the same nonzero plane wave solution and follow the above
procedure except that one has to choose the $\hat{m}^{(1)}$ matrix
of the following form
\begin{eqnarray}
\hat{m}^{(1)}=\left( {\begin{array}{*{20}c}
   m_1 & 1/n_1  \\
   n_1 & {1/m_1}  \\
\end{array}} \right)
\end{eqnarray}
\begin{equation}
m_1=\frac{k + i \lambda}{|c_i|}\hspace{0.2cm}and\hspace{0.2cm}
n_1=i,
\end{equation}
Then, equation.(\ref{darkgenone}) yields the dark soliton solution
for the corresponding integrable scalar NLS type equation.
\section{Applications}
\textbf{(i)     Coupled GP (CGP) equation}: \vspace{10pt}

The coupled GP equation representing a binary BEC comprising of
the hyperfine states of the same atomic species say, Rubidium
($^{87}$ Rb) in a transient harmonic trap can be written down (in
dimensionless form) as
\begin{equation}
i\frac{\partial\psi_{j}}{\partial t}+\frac{\partial^2
\psi_{j}}{\partial
x^2}+\Big[a(t)\sum_{k=1}^{2}b_{jk}|\psi_{k}|^2+v(x,t)\Big]\psi_j,
\label{eq:gp}
\end{equation}
In the above equation, $v(x,t)=\lambda(t)^2 x^2$ represents the
transient harmonic trap and $a(t)$ represents the temporal
scattering length (a(t) should be negative for defocussing
(attractive) and positive for focussing (repulsive) cases
respectively) and $\psi_j$, $j=1, 2$ describes the order parameter
of the condensates. The above integrable coupled GP equation has
also been investigated \cite{cgp-vr,cgp-sr} and the dynamics of
the vector BECs has been explored by constructing bright and dark
solitons.

The above equation.(\ref{eq:gp}) admits the following eigenvalue
problem,
\begin{eqnarray}
\phi_x=U \phi \notag\\
\phi_t=V \phi \label{evpgp}
\end{eqnarray}
where, $\phi=(\phi_1,\phi_2,\phi_3)^T$ and
\begin{eqnarray}
U &=& \left(%
\begin{array}{ccc}
i\zeta(t) & Q_{1} &  Q_{2}\\
-Q_{1}^*& -i\zeta(t) & 0 \\
-Q_{2}^*& 0 & -i\zeta(t) \\
\end{array}%
\right),
\end{eqnarray}
\begin{eqnarray}
V&=&\left(%
\begin{array}{ccc}
V_{11} & V_{12} & V_{13} \\
V_{21} & V_{22} & V_{23} \\
V_{31} & V_{32} & V_{33} \\
\end{array}%
\right),
\end{eqnarray}
where,
\begin{eqnarray}
V_{11}&=& -i \zeta(t)^{2}+ i \Gamma(t) x \zeta(t)
+\frac{i}{2}\gamma(t) A(t) Q_{1} Q_{1}^{\ast}
+\frac{i}{2}\gamma(t) A(t) Q_{2} Q_{2}^{\ast}\notag\\
V_{12}&=& [\Gamma(t) x - \zeta(t)] Q_{1} + \frac{1}{2} Q_{1x}\notag\\
V_{13}&=& [\Gamma(t) x - \zeta(t)] Q_{2} + \frac{1}{2} Q_{2x}\notag\\
V_{21}&=& -[\Gamma(t) x - \zeta(t)] Q_{1}^{\ast} + \frac{1}{2} Q_{1x}^{\ast}\notag\\
V_{22}&=& i \zeta(t)^{2}- i \Gamma(t) x \zeta(t)
-\frac{i}{2}\gamma(t) A(t) Q_{1} Q_{1}^{\ast}\notag\\
V_{23}&=& -\frac{i}{2} Q_{2} Q_{1}^{\ast}\notag\\
V_{31}&=& -[\Gamma(t) x - \zeta(t)] Q_{2}^{\ast} + \frac{1}{2} Q_{2x}^{\ast}\notag\\
V_{32}&=& -\frac{i}{2} Q_{1} Q_{2}^{\ast}\notag\\
V_{33}&=& i \zeta(t)^{2}- i \Gamma(t) x \zeta(t)
-\frac{i}{2}\gamma(t) A(t) Q_{2} Q_{2}^{\ast}\notag
\end{eqnarray}
with
\begin{eqnarray}
Q_1=\frac{1}{\sqrt{A(t)}} \psi_1(x,t) e^{i (-\Gamma(t) x^2)}, \notag\\
Q_2=\frac{1}{\sqrt{A(t)}} \psi_2(x,t) e^{i (-\Gamma(t) x^2)}.
\notag\\\label{eq:trans}
\end{eqnarray}
In the above nonisospectral eigenvalue problem, spectral parameter
$\zeta(t)$ obeys the following equation
\begin{equation}
\zeta(t)=\mu e^{-(\int \Gamma(t) dt)}\label{eq:mu}
\end{equation}
where $\mu$ is a hidden complex constant and $\Gamma(t)$ is an
arbitrary function of time and
\begin{equation}
\Gamma(t)=\frac{d}{dt}log A(t).\label{gamm}
\end{equation}
\begin{equation}
a(t)=\frac{1}{A(t)},\label{sl}
\end{equation}
\begin{equation}
\lambda(t)^{2}=\Gamma(t)^2-\Gamma'(t)\label{trap}
\end{equation}
\begin{figure}
\begin{center}
\includegraphics[scale=0.4]{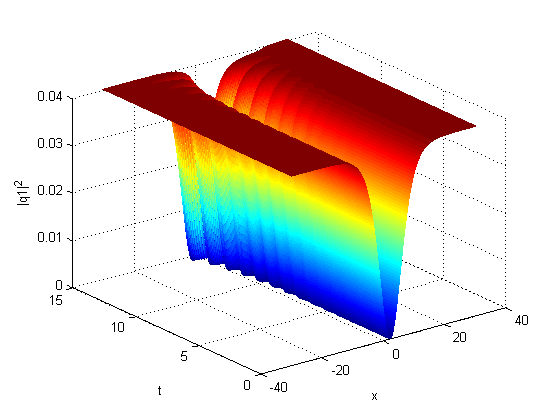}
\includegraphics[scale=0.4]{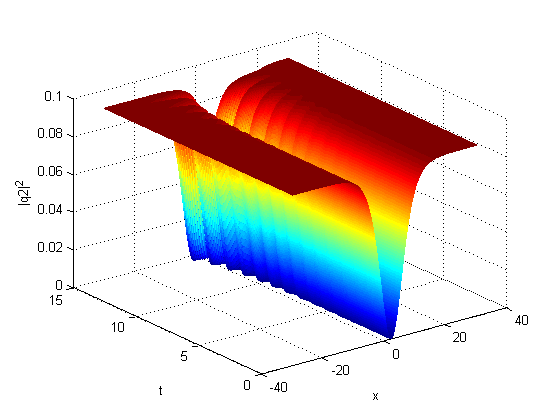}
\caption{The density profiles of the dark solitons  in the
presence of periodic modulated potential for the choice of
parameters $\Gamma(t)= cos(\omega t + \delta )e^{\sigma t}$,
$\omega$=6,$\delta$=0,$\sigma$=0.4,$\varepsilon_1^{(1)}$=0.3,
$\alpha_{10}$=0.1, $\beta_{10}$=0.3, $\chi_{1}$=0.1,
$\delta_{1}$=0.2,$a_{1}$=-$a_{2}$=1,$c_{1}$=$c_{2}$=1 }
\label{fig.1}
\end{center}
\end{figure}

\begin{figure}
\begin{center}
\includegraphics[scale=0.4]{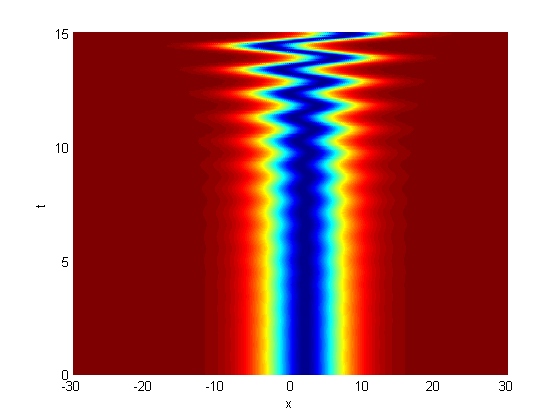}
\includegraphics[scale=0.4]{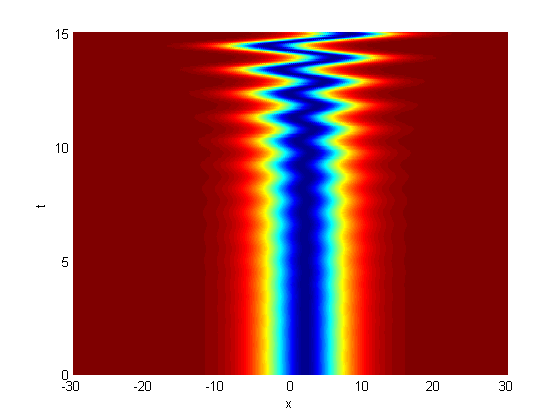}
\caption{Contour plots of fig.\ref{fig.1} exhibiting the beating
effect in the propagation dark solitons } \label{fig.2}
\end{center}
\end{figure}

\begin{figure}
\begin{center}
\includegraphics[scale=0.5]{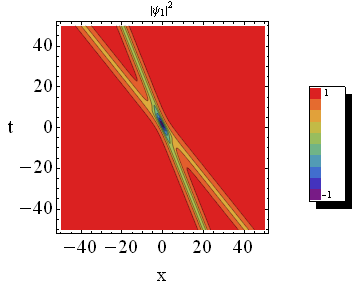}
\includegraphics[scale=0.5]{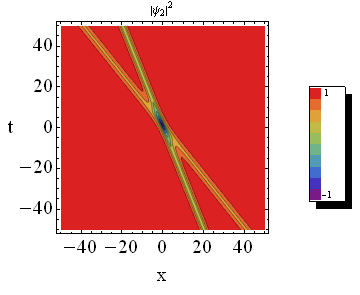}
\caption{Collisional dynamics of dark solitons for the choice of
parameters $a(t)= 0.1 $,$ \Gamma(t) = 0.1\times 10^-2 t$,
$\alpha_{1}=0.1$, $\alpha_{2}=0.25$, $\beta_{1}=0.3$,
$\beta_{2}=0.25$, $\delta_1 =0.1$, $\delta_2 = 0.2$, $\chi_1 =
0.3$, $\chi_2 = 0.4$, $\varepsilon_1^{(1)}=0.85i$,
$\varepsilon_1^{(2)}=0.5$ such that
$|\varepsilon_1^{(j)}|^2+|\varepsilon_2^{(j)}|^2 = 1, (j =
1,2)$,$a_1=-a_2=1,c_1=c_2=1$ } \label{fig.3}
\end{center}
\end{figure}

It is known \cite{cgp-sr,cgp-vr} that the coupled GP equation is
completely integrable only if the transient trap $\lambda(t)$ and
the scattering length $a(t)$ are connected by the following
equation (obtained by substituting eqn.(\ref{gamm}) and
eqn.(\ref{sl}) into eqn.(\ref{trap}))

\begin{equation}
-\frac{1}{2a(t)}\frac{d^2
a(t)}{dt^2}+\frac{1}{a^{2}(t)}\Big(\frac{d
a(t)}{dt}\Big)^{2}+2\lambda^2=0,\label{intcond}
\end{equation}

To generate dark vector solitons of the above GP eq.(\ref{eq:gp}),
we  choose the following non zero plane wave solution as the seed
\begin{equation}
q_i=c_{i} e^{[i(a_i x-(a_i^2/2+\sigma_l c_l^2)t)]} \label{nonzero}
\end{equation}
to obtain the vacuum eigenfunction
\begin{equation}
\Phi^{(0)} = \left(%
\begin{array}{ccc}
  \phi^{(0)}_{11} & 0 & 0 \\
  0 & \phi^{(0)}_{22} & 0 \\
  0 & 0 & \phi^{(0)}_{33} \\
\end{array}%
\right)exp(i(\mu x+(\zeta \mu
-\frac{1}{2}\zeta^2+\frac{1}{2}\mu^2+\sigma_l c_l^2)t))
\end{equation}
where
\begin{eqnarray}
\phi^{(0)}_{11}&=&e^{-2 i\zeta(t) x - 6i \zeta(t)^2 t},\notag\\
\phi^{(0)}_{22}&=&e^{i \zeta(t) x + 3i \zeta(t)^2 t},\notag\\
\phi^{(0)}_{33}&=&e^{i \zeta(t) x + 3i \zeta(t)^2 t},\notag
\end{eqnarray}
Employing the gauge transformation method, one obtains the dark
soliton solution for the coupled GP equation (\ref{eq:gp}) (for
the defocussing  case keeping the temporal scattering length
a(t)=-a(t)) of the following form
\begin{eqnarray}
\psi_1^{(1)} =\sqrt{\frac{1}{a(t)}} 2 c_{i}\varepsilon_1^{(1)}
\beta_1(t) tanh(\theta_1)e^{i(-\xi_1 + f(t) \frac{x^2}{2})},\label{onesolgp1}\\
\psi_2^{(1)} =\sqrt{\frac{1}{a(t)}} 2 c_{i}\varepsilon_2^{(1)}
\beta_1(t) tanh(\theta_1)e^{i(-\xi_1 + f(t)
\frac{x^2}{2})},\label{onesolgp2}
\end{eqnarray}
where
\begin{eqnarray}
\theta_1 &=& 2 \beta_1 x + 4\int \alpha_1 \beta_1 dt -2 \delta_1,\\
\xi_1 &=& 2 \alpha_1 x + 2\int(\alpha_1^2-\beta_1^2)dt+\Lambda
-2\chi_1,
\end{eqnarray}
and
\begin{equation}
\Lambda=e^{[i(a_{i} x-(a_{i}^2/2+\sigma_l c_{i}^{2})t)]}
\end{equation}
with $\alpha_1 = \alpha_{10} e^{[\int{\Gamma(t)}dt]} $,
$\beta_1=\beta_{10} e^{[\int{\Gamma(t)}dt]} $ while $\delta_1$ and
$\chi_1$ are arbitrary parameters and
$\varepsilon_1^{(1)}$,$\varepsilon_2^{(1)}$ are coupling
parameters connected by the relation
$|\varepsilon_1^{(j)}|^2+|\varepsilon_2^{(j)}|^2 = 1, (j = 1,2)$.
The dark solitons given by eqs.(\ref{onesolgp1}) and
(\ref{onesolgp2}) are identical to the one reported in
\cite{cgp-sr}. The density profile of dark solitons given by
eqs.(\ref{onesolgp1}) and (\ref{onesolgp2}) is shown in
fig.(\ref{fig.1}) while its contour plots shown in
fig.(\ref{fig.2}) display the beating effect of the dark solitons
during time evolution. It should be mentioned that this beating
effect arises due to the temporal nature of the harmonic trap
\cite{cgp-sr,snakeeff}. Eventhough the nature of the solitons
depends on the scattering length a(t) and trap frequency
$\lambda(t)$ (or $\Gamma(t)$) in accordance with the integrabilily
condition given by equation.(\ref{intcond}), the trap frequency
$\Gamma(t)$ predominates over the scattering length. As we have
chosen the trap frequency $\Gamma(t)$ to be a periodic wave with
exponentially varying amplitude ($\sigma
>0$), the amplitude of the soliton visibly
gets higher for large "t" as it is evident from figs.\ref{fig.1}
and fig.(\ref{fig.2}).This beating effect of dark solitons is
consistent with the experimental results discussed in
ref.\cite{experimental}. Our paper also gives a simple
experimental protocol to observe "beating effect" in the
collisional dynamics of dark solitons without the impact of
external thermal cloud.The collisional dynamics of two solitons is
shown in fig.(\ref{fig.3}). One can easily extend this modified
gauge transformation approach to construct multi dark soliton
solutions and study their collisional dynamics.

It should also be mentioned that the present integrable model does
not have the luxury of observing parametric resonance excitations
in dark solitons (BECs) by virtue of the constraint imposed by the
integrability  condition given by eq.(\ref{intcond}) as one may
not be able to choose the frequency of the trap $\Gamma(t)$ and
$a(t)$ desirably.


\vspace{10pt} \textbf{(ii) Coupled Nonlinear Schr\"odinger (CNLS)
equation} \vspace{10pt}

Under the dependent variable transformation
\begin{eqnarray}
\psi_{1}=\frac{1}{\sqrt{a(t)}l(t)}\Phi_{1}(X,T)exp(i[\Gamma(t)x^2])\\
\psi_{2}=\frac{1}{\sqrt{a(t)}l(t)}\Phi_{2}(X,T)exp(i[\Gamma(t)x^2])
\end{eqnarray}
where $\Phi_i$ is an arbitrary function with spatial and temporal
variables chosen as  $X=\frac{x}{l(t)}$ and $T=l^{-2}$ with
$\Gamma(t), \lambda(t)$ given by
\begin{equation}
\Gamma_{t}+4\Gamma^2+\lambda^2=0, l_t-4\Gamma l=0,
\end{equation}
eqn.(\ref{eq:gp}) can be reduced to the standard coupled nonlinear
Sch\"odinger equation in the normal GVD region of the following
form
\begin{eqnarray}
i\Phi_{1T}-\Phi_{1XX}+\nu(|\Phi_1|^2+|\Phi_2|^2)\Phi_{1}=0,\\
i\Phi_{2T}-\Phi_{2XX}+\nu(|\Phi_1|^2+|\Phi_2|^2)\Phi_{2}=0.
\end{eqnarray}
\begin{figure}
\begin{center}
\includegraphics[scale=0.425]{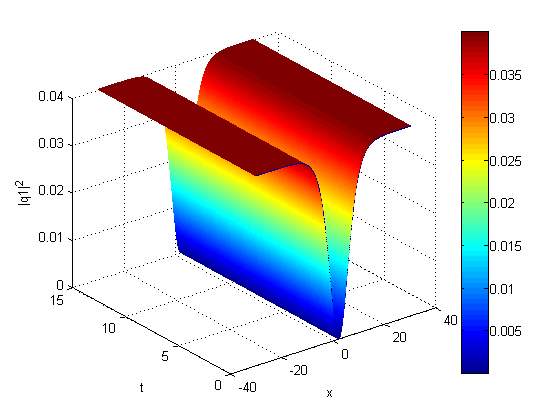}
\includegraphics[scale=0.425]{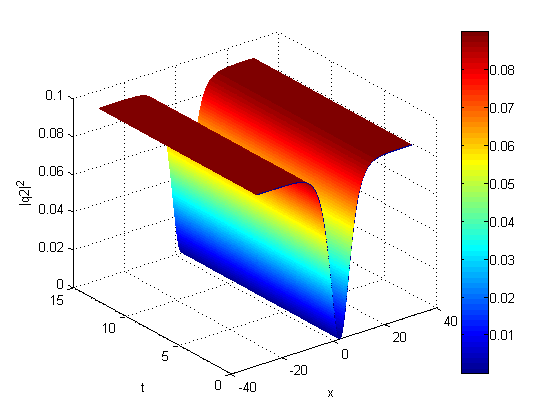}
\caption{Density profiles of  dark solitons  of coupled NLS
equation for the choice $\varepsilon_1^{(1)}=0.3$,
$\alpha_{10}=0.3, \beta_{10}=0.5, \chi_1=0.3,
\delta_1=0.5$,$a_1=-a_2=1,c_1=c_2=1$ }\label{fig.4}
\end{center}
\end{figure}

\begin{figure}
\begin{center}
\includegraphics[scale=0.425]{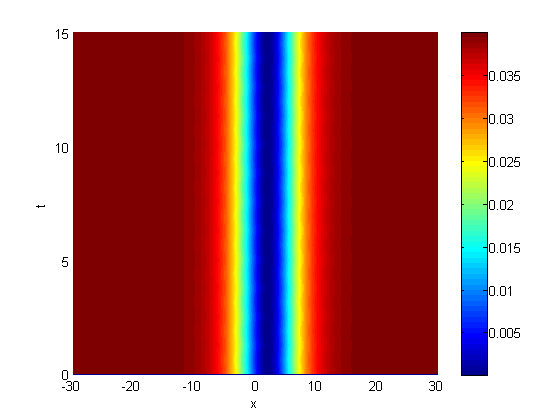}
\includegraphics[scale=0.425]{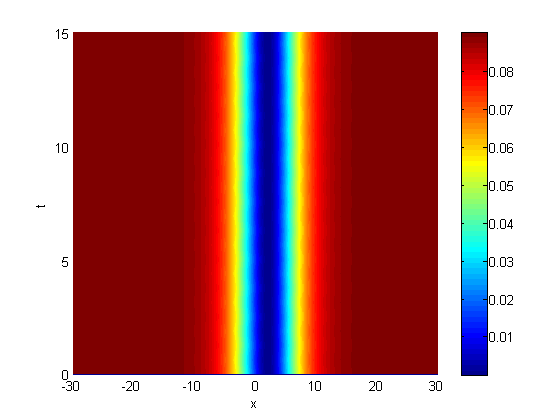}
\caption{contour plots of fig.(\ref{fig.4}) showing the trajectory
of dark solitons}\label{fig.5}
\end{center}
\end{figure}

\begin{figure}
\begin{center}
\includegraphics[scale=0.475]{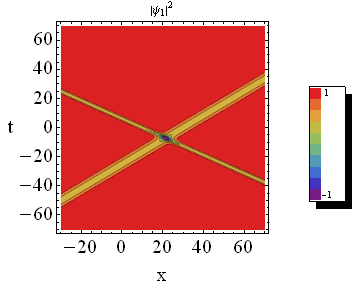}
\includegraphics[scale=0.475]{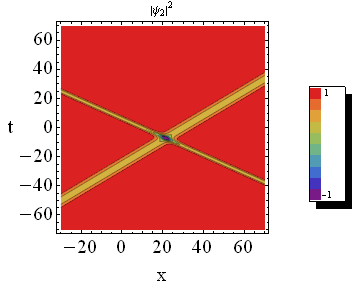}
\caption{Collisional dynamics of dark solitons of the coupled NLS
equation for the parametric choice $\alpha_1=0.1$,
$\alpha_2=0.25$, $\beta_{1}=0.3$, $\beta_{2}=0.2$, $\delta_1
=0.1$, $\delta_2 = 0.2$, $\chi_1 = 0.3$, $\chi_2 = 0.4$,
$\varepsilon_1^{(1)}=0.85i$, $\varepsilon_1^{(2)}=0.5$ such that
$|\varepsilon_1^{(j)}|^2+|\varepsilon_2^{(j)}|^2 = 1/\nu, (j =
1,2)$, $a_1=-a_2=1,c_1=c_2=1$}\label{fig.6}
\end{center}
\end{figure}


The above coupled nonlinear equation is nothing but the celebrated
integrable model proposed by Manakov \cite{svm}. The dark soliton
solution of the CNLS equations employing the gauge transformation
approach is of the following form (defocussing- defocussing
Manakov model for $\nu=-1$)
\begin{subequations}
\begin{eqnarray}
\psi_1^{(1)} = - c_1 \varepsilon_1^{(1)}
\beta_1 tanh(\theta_1)e^{i(-\xi_1)},\\
\psi_2^{(1)} = - c_2 \varepsilon_2^{(1)} \beta_1
tanh(\theta_1)e^{i(-\xi_1 )},
\end{eqnarray}\label{darkonesol.nls}
\end{subequations}
where
\begin{eqnarray}
\theta_1 &=& 2 \beta_1 x + 8 \alpha_1 \beta_1 t-2 \delta_1,\nonumber\\
 \xi_1 &=& 2 \alpha_1 x + 4(\alpha_1^2-\beta_1^2)t+\Lambda-2\chi_1,\nonumber
\end{eqnarray}
and
\begin{equation}
\Lambda=e^{[i(a_{i} x-(a_{i}^2/2+\sigma_l c_{i}^{2})t)]}
\end{equation}
subject to $|\varepsilon_1^{(j)}|^2+|\varepsilon_2^{(j)}|^2
=\frac{1}{\nu}, (j = 1,2)$ where $\varepsilon_{1,2}$ represent
coupling parameters. $\alpha_{j}=Re[\mu]$ and $\beta_{j}=Im[\mu]$
while $j=1,2$ are  the real constants and $\mu$ is the hidden
spectral parameter. The dark soliton given by
eqn.(\ref{darkonesol.nls}) is identical to the one reported in ref
\cite{RRK:ML:nls}. The density profile of dark solitons  and its
trajectory are shown in figs.(\ref{fig.4}) and (\ref{fig.5})
respectively. The collisional dynamics of dark solitons of the
Manakov model is displayed in fig.(\ref{fig.6}).

\section{Conclusion}
In this paper, we have formulated a simple algebraic approach by
harnessing the freedom in the celebrated gauge transformation
approach to construct dark solitons of coupled NLS type equations.
As an application, we have constructed the dark solitons of
coupled GP and coupled NLS equations and studied their properties.
It should be emphasized that the present approach is purely
algebraic and enables one to generate multi dark soliton solutions
from a trivial/nontrivial seed solution and it is quite superior
to other analytical techniques like Darboux transformation method
\cite{DT}, Hirota method \cite{RRK:ML:nls,Hirota}  and IST
\cite{IST}. As far the limitations of this approach are concerned,
it should be emphasized that this method generates only bright and
dark solitons for zero and non zero seeds respectively. We have
not yet exploited it to generate breathers, Ma solitons, rogue
waves amongst others.

\textbf{Acknowledgements:} PSV wishes to thank UGC for the
financial support. RR wishes to acknowledge the financial
assistance received from DST (Ref.No:SR/S2/HEP-26/2012) and UGC
(Ref.No:F.No 40-420/2011(SR) dated 4.July.2011). KP acknowledges
DST, NBHM, DST-FCT, IFCPAR and CSIR, Government of India for the
financial support through major projects.


\begin{thebibliography}{00}
\bibitem{krusal1965}
N.J.Zabusky, M.D.Kruskal, Phys. Rev. Lett. {\bf 15}, 240 (1965).
\bibitem{mkdv}
R. M. Miura, C. S. Gardner and M. D. Kruskal, J. Math. Phys. {\bf
9}, 1204 (1968).
\bibitem{sg}
Ryogo Hirota, J. Phys. Soc. Jpn. {\bf 33}, 1459 (1972).
\bibitem{nls}
V. E. Zhakharov and E. I. Schulman, Physica D {\bf 4},  270
(1982).
\bibitem{ablo}
A. T. Avelar,  D. Bazeia,  and W. B. Cardoso,  Phys. Rev. E {\bf
79}, 025602(R) (2009).\newline D J Frantzeskakis, J. Phys. A:
Math. Theor. {\bf 43} 213001 (2010)\newline L A Toikka and K-A
Suominen, Phys. Rev. A {\bf 87}, 043601 (2013).
\bibitem{hase1}
A Hasegawa and Y Kodama, \textit{Solitons in Optical
Communications} (Oxford University press, Oxford, 1995).
\bibitem{hase2}
A Hasegawa and F D Tappert, Appl.Phys. Lett. {\bf 23}, 142
(1973).
\bibitem{gpa}
G. P Agrawal, \textit{Nonlinear Fibre Optics} (Academic, New York,
2012).
\bibitem{darksoliton.nlo}
K. J. Blow and N. J. Doran, Phys. Lett. A {\bf 107}, 55  (1985).
\bibitem{PSV:PRE}
R. Radha, P. S. Vinayagam and K. Porsezian, Phys. Rev. E. {\bf
88}, 032903  (2013).

\bibitem{ofc}
M. F. Saleh and F. Biancalana, Phys. Rev. A  {\bf 87}, 043807
(2013).
\bibitem{jrt}
J. R. Taylor, \textit{Optical Solitons-Theory and Experiment}
(Cambridge University press, New York  1992).

\bibitem{dkr}
D. Krokel, N. J. Halas, G. Giuliani and D. Grischkowsky, Phys.
Rev. Lett. {\bf 60}, 29  (1988).
\bibitem{amw}
A. M. Weiner, J. P. Heritage, R. J. Hawkins, R. N. Thurston, E. M.
Kirschner, D. E. Leaird and W. J. Tomlinson, Phys. Rev. Lett. {\bf
61},  2445 (1988).
\bibitem{hblm}
Ryogo Hirota and Junkichi Satsuma, Prog. Theor. Phys. Supplement.
{\bf 59},  64 (1976).
\bibitem{hblm:2}
R. K. Bullough and P. J. Caudrey \textit{Solitons} (Springer,
Berlin, 1980) p. 157.
\bibitem{mateev}
V. B. Matveev, Phys. Lett. A {\bf 166},  205 (1992).
\bibitem{mateev:2}
V. B. Matveev and M. Salle, \textit{Darboux transformations and
Solitons} (Springer- Verlag, Berlin, 1991).
\bibitem{ist}
Clifford S.Gardner, John.M.Greene, Martin.D.Kruskal and Robert.M.
Miura, Phys. Rev. Lett. {\bf 19},  1095 (1967).
\bibitem{llchaw1991}
L.-L. Chau, J.C. Shaw, H.C. Yen, J. Math. Phys. {\bf 32(7)},
 1737 (1991).

\bibitem{svm}
S V Manakov, Sov. Phys. JETP. {\bf 38(2)},  248 (1974).


\bibitem{cgp-vr} V. Ramesh Kumar, R. Radha and M. Wadati,  Phys.Lett.A
\textbf{374},  3685 (2010).

\bibitem{cgp-sr} S. Rajendran, P. Muruganandam and M. Lakshmanan, J. Phys. B:
At. Mol. Opt. Phys. \textbf{42},  145307 (2009).

\bibitem{snakeeff} Vladimir N. Serkin and Akira Hasegawa, Phys.
Rev. Lett. {\bf 85}, 4502 (2000).

\bibitem{experimental}
C. Backer, S. Stellmer, Soltan-Panahi, S. Dorscher, M. Baumert, E.
M. Richter, J. Kronjager, K. Bongs, Sengstock and Klaus, Nature
Phys. {\bf 4} 496, (2008)\\
A. Weller, J. P. Ronzheimer,C. Gross, D. J. Frantzeskakis, G.
Theocharis, P. G. Kevrekidis, J. Esteve and M. K. Oberthaler, Phys
Rev L, {\bf 101} 130401 (2008)\\
S. Stellmer, C. Becker, P. Soltan-Panahi, E. M. Richter, S.
D¨orscher, M. Baumert, J. Kronj¨ager, K. Bongs  and K. Sengstock,
Phys Rev L, {\bf 101} 120406 (2008)


\bibitem{RRK:ML:nls}
R. Radhakrishnan and M. Lakshmanan, J. Phys. A:Math. Gen. {\bf
28}, 2683 (1995).

\bibitem{DT} Liming Ling, Li-Chen Zhao and Boling Guo,
arXiv:1309.1037v1[nlin.SI].

\bibitem{Hirota} A. Mahalingam and K. Porsezian, Phys. Rev. E. {\bf
64}, 046608 (2001).

\bibitem{IST} M. J. Ablowitz, G. Biondini and B. Prinari, J. Math.
Phys. {\bf 47}, 063508 (2006).

\end{thebibliography}
\end{document}